\documentclass{emulateapj}
\usepackage{amsmath}
\def \lp{\>\> .}
\def \lc{\>\> ,}

\def \c2{cm$^{-2}$}

\def \nh2{n_{H_2}}
\def \nh1{n_{HI}}

\def \nh3{NH$_3$}
\def \n2h{N$_2$H$^+$}

\def \hh{H$_2$}

\def \ox18{$^{16}$O$^{18}$O}

\def \cp{C$^+$}
\def\CII{C{\sc ii}}

\def \HI{H{\sc i}}

\def\n3{$n_u$}
\def\n1{$n_l$}

\def \mic{$\mu$m}

\def \be{\begin{equation}}
\def \ee{   \end{equation}}
\def \bf {\begin{figure}}
\def \ef {   \end{figure}}
\def \bc{\begin{center}}
\def \ec{\begin{center}}

\def \lc{\>\> ,}
\def \lp{\>\> .}

\begin{document}
\title{ \CII\ in the Interstellar Medium:  Collisional Excitation by \hh\  Revisited}
\setcounter{footnote}{0}

\author{Laurent Wiesenfeld\altaffilmark{1,2} and Paul F. Goldsmith\altaffilmark{2}}

\altaffiltext{1} {UJF-Grenoble 1/CNRS-INSU, Institut de Plan\'{e}tologie et d'Astrophysique de Grenoble (IPAG), UMR 5274, Grenoble, France}
\altaffiltext{2}{Jet Propulsion Laboratory, California Institute of Technology, Pasadena CA}


\begin{abstract}

\cp\ is a critical constituent of many regions of the interstellar medium, as it can be a major reservoir of carbon and, under a wide range of conditions, the dominant gas coolant.
Emission from its 158 \mic\ fine structure line is used to trace the structure of photon dominated regions in the Milky Way and is often employed as a measure of the star formation rate in external galaxies.
Under most conditions, the emission from the single [\CII] line is proportional to the collisional excitation rate coefficient.
We here used improved calculations of the deexcitation rate of [\CII] by collisions with \hh\ to calculate more accurate expressions for interstellar \cp\ fine structure emission, its critical density, and its cooling rate.   
The collision rates in the new quantum calculation are $\sim$ 25\% larger than those previously available, and narrow the difference between rates for excitation by atomic and molecular hydrogen.
This results in [\CII] excitation being quasi-independent of the molecular fraction and thus dependent only on the total hydrogen particle density.  
A convenient expression for the cooling rate at temperatures between 20 K and 400 K, assuming an LTE  H$_2$ ortho to para ration is $\Lambda ({\rm LTE~OPR}) = \left(11.5 + 4.0\,e^{-100\,\mathrm K/T^{\rm kin}}\right)\;e^{-91.25\,\mathrm K/T^{\rm kin}}\,n ({\rm C}^{+})\,n({\rm H}_2)\times 10^{-24}\;{\rm ergs}~{\rm cm}^{-3}~{\rm s}^{-1}$.  
The present work should allow more accurate and convenient analysis of the [\CII] line emission and its cooling.

\end{abstract}


\section{INTRODUCTION}

Ionized carbon is distributed throughout a large fraction of the interstellar medium, from ionized regions to dense clouds that are largely molecular.
Due to its high abundance and equivalent temperature of $\simeq$ 90 K, the 158 \mic\  ($=63.395\;\mathrm{cm^{-1}}$) [\CII] fine structure transition plays a particularly important role in cooling the warm neutral medium (WNM) and contributes to the transformation of this gas 
into the cooler, denser, cool neutral medium (CNM).  
This spectral line is an excellent tracer of the ``CO-dark molecular gas'' \citep{langer2010, pineda2013}, in which hydrogen is molecular, but carbon is largely not in the form of CO, with the result that neither that molecule nor \HI\ trace this component of the interstellar medium \citep{wolfire2010}.
Neutral carbon is also in principle a tracer of the ``CO-dark molecular gas", but as indicated by the recent results of \citet{shimajiri2013}, the 609 \mic\ [CI] fine structure line traces the bulk of molecular cloud material as well. 

Due in part to its relatively large intensity (up to 1\% of the far--infrared luminosity of galaxies), [\CII] is a widely used -- if still incompletely understood -- tracer of star formation \citep{stacey2010}.
The above roles and uses of the [\CII] fine structure line are all dependent on the collisional excitation of the 
transition.
Since the excitation is generally subthermal, the emergent intensity is dependent on the collision rate 
coefficient and the colliding partner density.
In more diffuse regions, the excitation is by collisions with electrons and atomic hydrogen, but in denser, more shielded regions, collisions with molecular hydrogen are dominant.
In this paper we reexamine the rate coefficients for \cp--\hh\ collisions, in particular the recent calculations by 
\citet{LI13} (LI13).
In Section \ref{sec:potential}, we discuss the recent calculation and use the results to derive total [\CII] deexcitation rate coefficients  for collisions with \hh.
In Section \ref{sec:discussion} we discuss the implications of the new rates for collisional excitation, the critical density, and \cp\ cooling.

\section{POTENTIAL AND RATES}\label{sec:potential}

\subsection{The \cp - H$_2$ interaction energy}

The fine structure (de-)excitation collisional rates of \cp\ have been computed in several investigations found in the literature.
Collisions with atomic H have been treated quite comprehensively by \citet{barinovs05}, and   collisions with $e^-$\ by \citet{wilson02}. 
The collisions of \cp\ with \hh\ were first computed by \citet{chu1975}, but in considerably more detail in the work of  \citet{flower77a}, \citet[][hereafter FL77b]{flower77b}, and \citet{flower88} with some  approximations in the potential energy surface. 
Very  recently, LI13 recalculated both the $\mathrm{C^+ - H_2}$ interaction energy and the fine structure 
(de-)excitation cross section and rates.

In the LI13 paper, the interaction energy was computed with much greater precision than in the previous studies, thanks to 
the availability of more reliable and more complete quantum chemistry codes (MOLPRO code, \citep{molpro}) as well as much 
larger computing capacity. 
Large electronic basis sets were employed to describe as precisely as possible the molecular wave-functions. 
\cp\ being an open-shell atomic ion, care was taken to allow for low energy virtual electronic excitations. Hence, the interaction energy was computed using specific multireference configuration interaction (MRCI) methods, with the 2s and 2p valence electrons of \cp\ and the two electrons of H$_2$ being active. 
All relevant quantum chemical details may be found in LI13, section II.  

Since the $\mathrm{C^+ - H_2}$ interaction is not strongly dependent on $\theta$, where $\theta$ describes the \hh\ interatomic axis orientation with respect to the ${\mathrm C^+}- M$ axis ($M$, center of mass of \hh),  LI13 sampled only 5 values of $ \theta$, thereby limiting the tensorial expansion of the potential energy in $\theta$ to $l = 0,2$ ($l$ odd is forbidden as \hh\ is homonuclear; see equation 1 in LI13 or FL77b. 
Because the unpaired electron is \cp\ is in the 2p orbital, the total expansion of the interaction potential is limited to 5 terms, which describe the full interaction in the aforementioned  approximation.  

\begin{figure}[htbp]
\begin{center}
\includegraphics[width=0.52\textwidth]{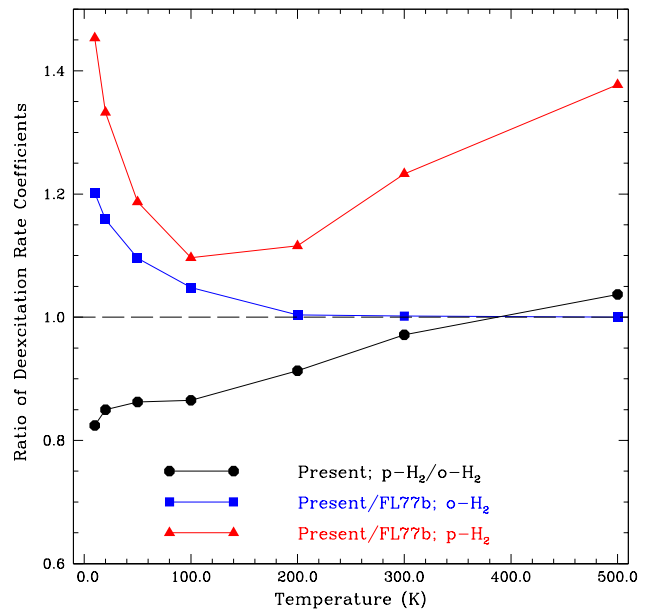}
\caption{ Ratio of the present R$_{ul}$ deexcitation rate coefficients (Table \ref{tab:rates}) relative to the \citet{flower77b} rate coefficients for para--\hh\ (triangles), the same ratio for ortho--\hh\ (squares), and the ratio of present rate coefficients for para--\hh\ relative to those for ortho--\hh.}
\label{fig:ratios}
\end{center}
\end{figure}

If we compare the LI13 and the FL77b results, the depths of the isotropic and anisotropic potentials are very different. 
Both isotropic and aniso\-tropic terms of the LI13 potential are  more attractive, which is an indication of the much better approximation  of the electronic wave-functions employed in LI13. 
It might thus appear that changes in the cross sections and rates for the fine structure transition for \cp ($\mathrm{^2P_{1/2}\leftrightarrow ^2P_{3/2}}$) (the [\CII] transition) would reflect this vastly different potential energy surface. 
It is remarkable that results of LI13 and  Figure~\ref{fig:ratios} here show the contrary. 
Cross sections are evidently affected, but by no more than a factor of 50\%. 
This result is in strong contrast to many other calculations. 
When examining rotational (de-)excitation rates, even a modest change of the interaction 
potential yields very different rates, with factors of 2-10 being not uncommon. 
The case of H$_2$O -- \hh\ collisions is discussed in \citet{dubernet2006}, while very small differences  in the CO--\hh\ interaction energies have non--negligible effects on calculated rates and cross sections \citep{yang2010}.

Here the situation  is very different.  
The fine structure changing rates have very little sensitivity to the depth of the potential, but depend primarily on the fine structure constant $\Delta E_{1/2 - 3/2} = 63.395\,\mathrm{cm^{-1}}$ and on the difference in energy of the two potential sheets $^2\Sigma$ and $^2\Pi$ (or $^2$A and  $^2\mathrm B_{1,2}$), as a close examination of formula (2) in \citet{launay77} shows. 
Indeed, these potential sheets describe the two polarizations of the electronic angular momentum 
(here $l=1$) with respect to the $\mathrm C^+ - M$ axis. A change of this polarization (from parallel to perpendicular) 
changes the value of the scalar product $\mathbf{l\cdot s}$ ($\mathbf s$, electronic spin), and hence induces a fine-structure transition (see 
\citet{mies73} for a very lucid exposition). 
One is entitled to view this energy difference in the potential sheets as a good representation of the collision-induced change in the fine-structure state of \cp.

Even with the more elaborate treatment of the \hh--\cp\ collision compared to previous treatments and the greater computer power available, it must be recognized that approximations have still been made.  For this reason, we estimate that an uncertainty of $\pm$ 20\% must be associated with the individual collision rate coefficients presented by LI13.

\subsection{\label{sec:rates} Rates}

Detailed rate coefficients for collisions with ortho-- and para--\hh\ were presented in LI13. 
In order for these rates to be usable for astrophysical applications, it is  necessary to average those rates over the \hh\ rotational populations, both for para-\hh\ and ortho-\hh. 
We assume here that the \hh\ molecule is in LTE for the ortho states ($J=1,3,\ldots$) and the para states ($J=0,2,\ldots$), separately. 
This assumption is discussed further in Section \ref{sec:LTE}.

For temperatures at which \cp\ emission and cooling are important ($T^{\mathrm{kin}} \geq 50\,\mathrm K$), ortho--to--para conversion is likely to be very slow due to short residence times on grain surfaces \citep{lebourlot1999}.  
Proton exchange through collisions with H$^+$ and H$_3^+$ are likely the dominant pathway for interconversion of the two spin modifications of \hh\ \citep{honvault2011}.
The moderately large rate coefficient of $\sim 10^{-10}\rm\, cm^{3}\,s^{-1}$ combined with low density of the relevant ions \citep[e.g.][]{wilgenbus2000} yields characteristic timescales of 10$^6$ to 10$^7$ years, which is comparable to or longer than the ages of these regions.
Thus, we do not suppose {\it a priori} equilibrium between ortho- and para-\hh, considering them as separate species.

We take a value of $B(\mathrm H_2)=60.853\,\mathrm{cm}^{-1}$, its average value for $v=0$. The $J=2$ population for para-
\hh\ becomes appreciable for  $T\geq 200\,\rm K$. 
Since the $J(\mathrm H_2)=2 \leftrightarrow 2$ rates are larger than the $J(\mathrm H_2)=0 \leftrightarrow 0$ rates, and that the $J(\mathrm H_2)=2 \leftrightarrow 0$ rates are by no means negligible, the para-H$_2$ rates become comparable to (or even slightly larger than) the ortho-H$_2$ rates at sufficiently high temperature. 
Note that for the temperature range considered here, the $J(\mathrm H_2) =3$ population remains very small.

The ortho-\hh\ and para-\hh\ rates are presented in Table \ref{tab:rates}, and a comparison with the results derived from the cooling curve presented by \citet{flower77b} in Figure~\ref{fig:ratios}.
Because of the importance of the $J(\mathrm H_2) =2$ population, the difference between the rates of ortho- and para-\hh\ becomes very small at higher temperatures.
The ortho to para ratio (OPR) of \hh\ is  not a very important parameter here, contrary to the situation for many rotational excitation rates \citep{troscompt2009,wiesenfeld2013}. 
While for rotational excitation, there is a strong dependance on the quadrupole interaction of \hh\  (averaged to zero for para-H$_2$, $J=0$), these terms are  definitely not the dominant ones in the process of fine structure excitation considered here.

It is of interest to compare the various rates at our disposal for the \cp\ ion. 
The deexcitation rate coefficients for H \citep{barinovs05} increase slowly, from $\sim 6$ to $\sim 8\, \times \, 10^{-10}\;\mathrm{cm^{3}\,s^{-1}}$, as the temperature increases from 10~K to 100~K.
The deexcitation rate coefficients for \hh\ (Table \ref{tab:rates}) increase somewhat less rapidly as a function of temperature.

\begin{figure}
\begin{center}
\includegraphics [width=0.52\textwidth]{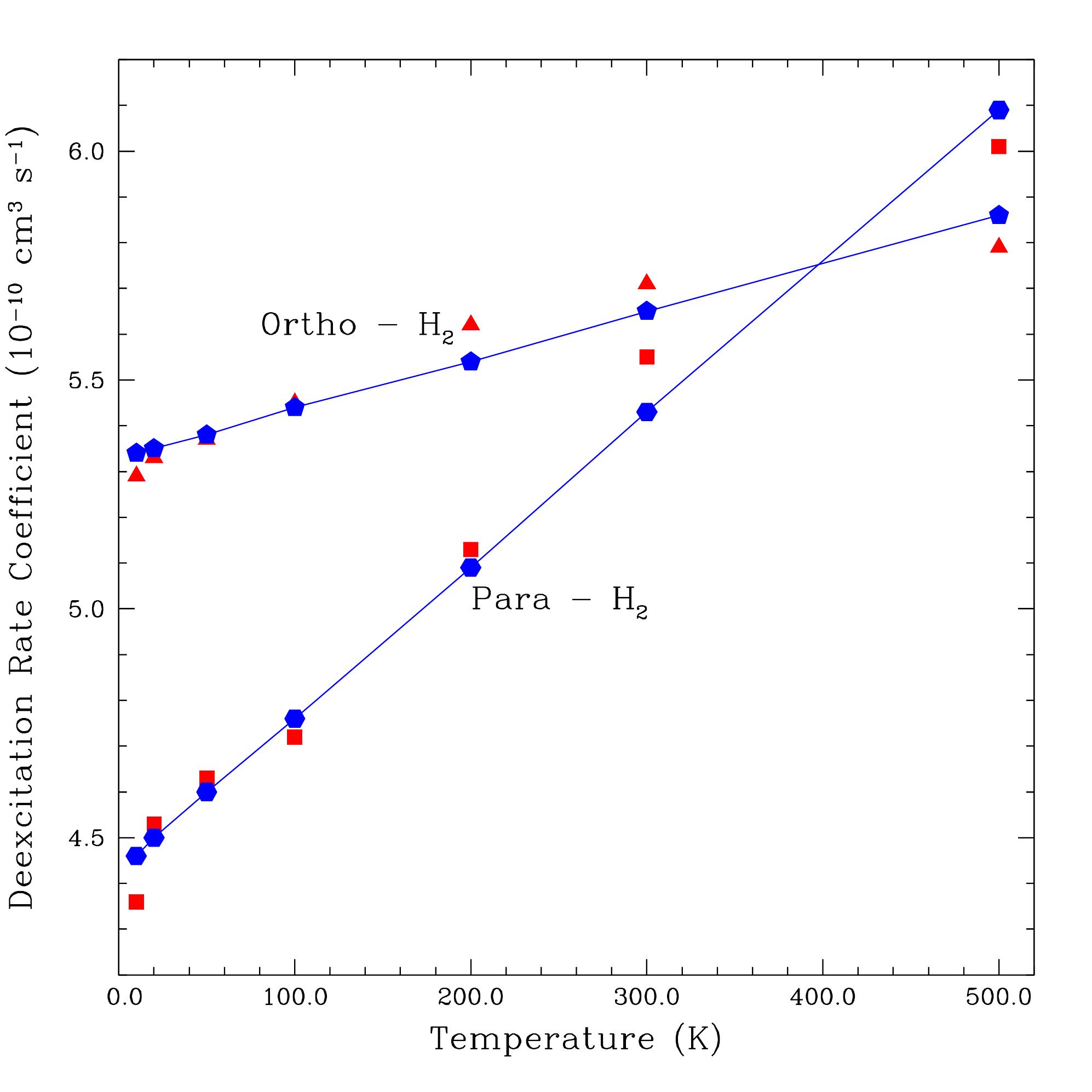}
\caption{\label{ortho_para_rates}  Collisional deexcitation rate coefficients R$_{ul}$ for the \cp\ fine structure transition by ortho--\hh\ and para--\hh\, as a function of temperature.
The rates from LI13 for para--\hh\ are indicated by squares and for ortho--\hh\ by triangles.  The least squares fit results for each spin modification are indicated by the symbols connected by solid lines.  The least squares fit results (equations \ref{lsf_para} and \ref{lsf_ortho}, respectively) are satisfactory for temperatures between $\simeq$ 20 K and $\simeq$ 400 K.
}
\end{center}
\end{figure}

In Figure \ref{ortho_para_rates} we show the rate coefficients from LI13, along with linear least-squares fit made for each spin modification separately.  
We find the least-squares results, valid over the temperature range betweeen $\simeq$ 20 K and $\simeq$ 400 K, to be

\begin{equation}
\label{lsf_para}
R_{ul}({\rm p-H_2}) = \left(4.43 + 0.33\,\frac{T^{\rm kin}}{100~ {\rm K}}\right)\times10^{-10}\;{\rm cm}^3\,{\rm s}^{-1} \lc
\end{equation}
and
\begin{equation}
\label{lsf_ortho}
R_{ul}({\rm o-H_2}) =\left(5.33 + 0.11\,\frac{T^{\rm kin}}{100~ {\rm K}}\right)\times10^{-10}\;{\rm cm}^3\,{\rm s}^{-1} \lp
\end{equation}

For any fixed ortho to para ratio, the fit for the collisional deexciation rate coefficient is a linear interpolation of the two expressions above, and in particular,
\begin{equation}
\label{deex_OPR1}
R_{ul}({\rm OPR} = 1) = \left(4.9 + 0.22\,\frac{T^{\rm kin}}{100~ {\rm K}}\right)\times10^{-10}\;{\rm cm}^3\,{\rm s}^{-1} \lp
\end{equation}
The result is that (considering for example an \hh\ OPR equal to 1), the ratio of molecular to atomic rates decreases from $\sim 0.9$  for $T \leq 20\,\rm K$ to $\sim 0.7$ for $T \geq 100\,\rm K$.  
This ratio is somewhat larger than the 0.5 value adopted by \cite{goldsmith2012}. 
The collision rate coefficients with $e^-$ are much larger, of the order of  $4\times 10^{-7}\,\mathrm{cm^{3}\,s^{-1}}$\citep{wilson02}. 
For regions in which hydrogen is not ionized, the excitation by electrons can be neglected, while in HII regions electron excitation is dominant.
Hence, the collisional excitation (and the critical density, see Section \ref{sec:ex_cd}) of \cp, except in regions in which hydrogen is ionized, depends essentially on the total density of hydrogen, whether it be in atomic or molecular form, not on the densities of H, ortho-\hh\ or para-\hh\ individually.

\section{DISCUSSION}
\label{sec:discussion} 

\subsection{LTE}
\label{sec:LTE}
In Table \ref{tab:rates}, we give rates for collisions with H$_2$ assuming that for each separate spin modification, para-H$_2$ and ortho-H$_2$, the different rotational levels are in local thermodynamic equilibrium (LTE), as a result of $\mathrm{H_2-H_2}$ inelastic collisions. 
However, we consider no spin exchange, and the H$_2$ OPR is arbitrary.

It is of interest to examine the assumption of LTE because the timescales to establish LTE for H$_2$ are relatively long.
For para-H$_2$, the $J=2$ level is at 510 K above ground state. 
For LTE conditions, we would have $n\left[\mathrm H_2 (J=2)\right]/n\left[\mathrm H_2(J=0)\right]\simeq 0.39$ at 200~K. 
The deexcitation rate $R_{2,0}\left[\mathrm H_2 (J=2 \rightarrow 0)\right]$ has been computed a number of times in the literature, with the most recent and complete computations being those of \citet{lee08}. 
For collisions with para-H$_2$, the $R_{2,0}$ rate at 100~K is $5.5\times 10^{-13}$ $\mathrm{cm^3\,s^{-1}}$.  
With a spontaneous emission rate $A=2.9\times10^{-11}\,\mathrm{s^{-1}}$, the critical density for the 
$\mathrm H_2 (J=2 \rightarrow 0)$ transition is $n_{\rm cr}(\mathrm H_2) \simeq 50 \,\mathrm{cm^{-3}}$. 
Choosing whether the rate to be used for $\mathrm C^+$--para-H$_2$ collisions is the LTE para--H$_2$ rate or else the para--H$_2$ $J=0$ rate thus depends on the molecular hydrogen density.

 For ortho--H$_2$, because of the high energy of the $J=3$ level, ($E= 845\,\rm K$ above the $J=1$ level), the relevance of this level is minimal at $T \leq 500\,\mathrm K$, with an LTE population of only a few percent. 
 All the more, because with the larger spontaneous decay rate for the 3$\rightarrow$1 transition,  $4.8\,10^{-10}\,\mathrm{s^{-1}}$, 
and $R_{3,1}$ collisional deexcitation rate coefficient at 100~K =  $8\times10^{-13}$  $\mathrm{cm^3\,s^{-1}}$,
the critical density is higher: $n_{\rm cr}(\mathrm H_2)\simeq 600\,\mathrm{cm^{-3}}$ . 
Both effects suggest that neglecting the $J= 3$ rotational level is acceptable in many situations, but not in the warmest portions of dense PDR regions.
 
Assuming  that the rotational levels of \hh\  can be considered to be in LTE at the kinetic temperature, and if the spin states are populated according to their statistical weights,  ${\rm OPR} = 9\,\exp\left(-170.5\,\mathrm K/T^{\rm kin}\right)$ for ${T^{\rm kin} \leq 100\, {\rm K}}$ (that is, when the population of levels with $J>1$ are negligible).  
The ``transition'' from para--\hh\ at low temperatures to ortho--\hh\ at high temperatures results in a less linear behavior for the total rate coefficient  than those of the individual spin modifications shown in Figure \ref{ortho_para_rates}.
A satisfactory fit over the temperature range 20 K $\leq \mathit T\rm{^{kin}} \leq$ 400 K is given by the expression
\begin{equation}
\begin{split}
\label{OP_simple}
R_{ul}({\rm LTE~OPR}) = (4.55 + 1.6\,e^{-(100~{\rm K/\mathit T\rm{^{kin}})}}) \\
\times10^{-10}\;{\rm cm}^{3}\,{\rm s}^{-1} \lp
\end{split}
\end{equation}

\subsection{Excitation and [CII] Critical Density}
\label{sec:ex_cd}

Collisional excitation of the [\CII] fine structure line is discussed in some detail by \citet{goldsmith2012}.  
The present results (e.g.  Equations \ref{lsf_para} - \ref{OP_simple}) can be multiplied by the density of \hh\ and used in any of the expressions involving the upper to lower state collision rate, $C_{ul}$ (s$^{-1}$).   
As shown in Table \ref{rates} and Figures \ref{ortho_para_rates} and \ref{fig:ratios}, the [\CII] collisional deexcitation rate coefficients for the two spin modifications are slightly different.
Over the range of temperature for which the calculations are applicable and for which [\CII] emission is likely to be significant, we can use equation \ref{OP_simple}, which (with the A-coefficent 2.3$\times$10$^{-6}$ s$^{-1}$) yields for collisions with \hh\  at 100 K, $n_{\rm cr}$ = 4.5$\times$10$^{3}$ cm$^{-3}$.  
As a result of the slightly larger deexciation rate coefficient, this is $\sim$ 25\% lower than the value given by \citet{goldsmith2012} at the same temperature.

In the optically thin subthermal limit the intensity of the [\CII] line (or the antenna temperature it produces) is proportional to the column density of \cp\ multiplied by the collisional excitation rate \citep[see equation 30 in ][]{goldsmith2012}.
For analysis of emission from a PDR, for example, the larger collisional rate coefficients calculated here imply a correspondingly lower \cp\ column density or \hh\ volume density.  

\subsection{[\CII] Cooling}
\label{sec:  cooling}

The [\CII] fine structure line is expected theoretically \citep{hollenbach1999}, and found observationally \citep[e.g.][]{bernard2012}  to be a major coolant of the warm ISM including diffuse atomic and molecular clouds and photon dominated regions.  
[\CII] cooling is discussed in some detail in Section 7 of \citet{goldsmith2012}.  
The expressions given in Sections \ref{sec:rates} and \ref{sec:LTE} above for the deexcitation rate coefficients can be used in any of the expressions for the cooling rate with specified \hh\ ortho to para ratio (OPR).
A single expression for optically thin, subthermal [\CII] cooling in a region with molecular hydrogen having OPR equal to unity (which is not far from that found observationally by \citet{neufeld1998}) is obtained by substituting equation \ref{deex_OPR1} into the general expression (with $\Delta E_{lu}$ being the [\CII] transition energy):
\begin{equation}
\Lambda = {R_{lu}\,n({\rm C}^{+})\,n({\rm H}_2)}\,\Delta E_{lu} \lc
\end{equation}
which yields
\begin{equation}
\begin{split}
\Lambda({\rm OPR} = 1) = \left(12.3 + 0.55\,\frac{T^{\rm kin}}{100\,\mathrm K}\right)\;e^{{-91.25\,\mathrm K/T^{\rm kin}}} \\
\times \, n ({\rm C}^{+})\,n({\rm H}_2)\times 10^{-24}\;{\rm ergs}~{\rm cm}^{-3}~{\rm s}^{-1} \lp
\end{split}
\end{equation}

 For a temperature--dependent OPR and thus [CII] deexcitation rate (equation \ref{OP_simple}) the cooling rate per unit volume is

\begin{equation}
\label{lamda_opr1}
\begin{split}
\Lambda ({\rm LTE~OPR}) = \left(11.5 + 4.0\,e^{-100\,\mathrm K/T^{\rm kin}}\right)\;e^{-91.25\,\mathrm K/T^{\rm kin}} \\\
\times \, n ({\rm C}^{+})\,n({\rm H}_2)\times 10^{-24}\;{\rm ergs}~{\rm cm}^{-3}~{\rm s}^{-1} \lp
\end{split}
\end{equation}
This cooling rate, valid over the range $20\,\mathrm K\leq T^{\rm kin}\leq 400\,\mathrm K$, combined with that for excitation by atomic hydrogen given by  \cite{barinovs05}, should allow improved accuracy in analyzing [\CII] cooling and deriving ISM properties from observations of this fine structure transition \footnote{The cooling at lower temperatures can be calculated from collisional deexcitation rate coefficients given in Table \ref{rates}.}.

\section{SUMMARY}
We discuss recently-published improved rate coefficients for collisional deexcitation of the \cp\ fine structure line by ortho-- and para--\hh\ by \citet{LI13}.
We fit the temperature dependence of these rates to derive rates of collisional deexcitation of [CII] by \hh\ for various values of the ortho--to--para ratio, including that expected in LTE.  
We report the resulting changes in the collisional excitation, critical density, and cooling rate for \cp\ in regions in which the hydrogen is primarily molecular.  
We find a critical density for \hh\ collisions with \cp, which is lower than that previously available by $\sim$ 25\%, although this change is omparable to the uncertainties in the collision rate coefficients.
Our results reduce the \cp\ column density derived for an assumed \hh\ density by this factor and result in an increase of the [\CII] cooling rate by a similar factor.
While differing only modestly from previous rates, the new results should allow more accurate analysis of the 158 \mic\ \cp\ fine structure line and of its effect on the structure of interstellar clouds.

\section*{Acknowledgements}

This research was carried out in part at the Jet Propulsion Laboratory, which is operated by the California Institute of Technology under contract with the National Aeronautics and Space Administration. 
We thank Bill Langer for a careful reading of the paper and suggestions that improved it, as well as F. Lique, T. Stoecklin, and their coworkers, for useful discussions.
We appreciate a number of useful suggestions for improving the paper from the anonymous referee.
 LW thanks the Agence Nationale de la Recherche, contract ANR-12-BS05-0011-01 (HYDRIDES), the CNES (through Herschel Key Project CHESS), and JPL for their support of this work.

\begin{table}[htdp]
\caption{\label{rates} Rate coefficients (in units of $10^{-10}\mathrm{\,cm^3\,s^{-1}}$) for \hh\ collisional deexcitation  of the \cp\ 158 \mic\ transition as a function of temperature. The original rates are from LI13. \label{tab:rates}}
\begin{center}
\begin{tabular}{lrrrrrrr}
\hline\hline
Temperature  (K) & 10   & 20     &  50   & 100  & 200   &  300  & 500\\
\hline
Spin Modification\\

Para-\hh              & 4.36 & 4.53 & 4.63 & 4.72  & 5.13 & 5.55  &  6.01\\
Ortho-\hh            & 5.29 & 5.33 & 5.37 & 5.45  & 5.62 & 5.71 &   5.79 \\
\hline\hline
\end{tabular}
\end{center}
\end{table}

\end{document}